\documentclass[12pt]{iopart}
\usepackage{iopams}
\usepackage{setstack}
\usepackage{graphicx}
\usepackage{amsthm}

\newcommand{\be}{\begin{equation}}
\newcommand{\ee}{\end{equation}}
\newcommand{\bea}{\begin{eqnarray}}
\newcommand{\eea}{\end{eqnarray}}
\newcommand{\nn}{\nonumber}

\begin{document}

\title[Degeneracy, marginal 
stability and extremality of black hole horizons]{A note on degeneracy, marginal 
stability and extremality of black hole horizons}

\author{Jos\'e Luis Jaramillo}

\address{Max-Planck-Institut f{\"u}r Gravitationsphysik, Albert Einstein
Institut, Potsdam, Germany}
\begin{abstract}

Given a stationary axisymmetric black hole horizon admitting a section
characterised as a strictly future stable marginally outer trapped surface, we extend
the equivalence between the notions of horizon degeneracy and marginal stability 
to the fulfillment, under the dominant energy condition, 
of the $A=8\pi |J|$ geometric relation between the area $A$ and the angular momentum
$J$ of a horizon section.

\end{abstract}

\maketitle

We explore the relation among marginal stability, degeneracy
and rigidity in the $A\geq 8\pi |J|$ inequality for axisymmetric 
isolated horizons, in particular Killing horizons, revisiting certain known
(but disperse) results and filling some gaps in the literature.

Let $({\cal H}, [\ell^a])$ be an isolated horizon (IH) \cite{Ashtekar:2001jb}, 
with ${\cal H}=\mathbb{R}\times S^2$ and $[\ell^a]$ the equivalence 
class of null normals leaving invariant the horizon geometry, equivalent 
under constant rescalings. Let $({\cal H}, [\ell^a])$ be embedded
in a spacetime $({\cal M}, g_{ab})$ with Levi-Civita connection $\nabla_a$.
Let ${\cal S}$ be a spacelike section of ${\cal H}$
with induced metric $q_{ab}$, Levi-Civita connection $D_a$, Laplacian ${}^2\!\Delta$, 
Ricci scalar ${}^2\!R$ and volume element $\epsilon_{ab}$ 
($dS$ will denote the area measure). 
Let us choose a (future-oriented) representative $\ell^a$ and a null vector $k^a$, 
normalised as $\ell^a k_a = -1$, spanning the normal plane $T^\perp {\cal S}$. 
The expansion associated with a vector $v^a$ normal to ${\cal S}$
is  $\theta^{(v)} = q^{ab}\nabla_av_b$. 
In particular $\theta^{(\ell)}=0$ on any section ${\cal S}$, namely
a marginally outer trapped surface (MOTS). 
The normal fundamental form $\Omega^{(\ell)}_a$  and the 
non-affinity coefficient $\kappa^{(\ell)}$ (constant on an IH  \cite{Ashtekar:2001jb})
associated with $\ell^a$ are
\bea
\label{e:omega_kappa}
\Omega^{(\ell)}_a = - k^c {q^d}_a \nabla_d \ell_c \qquad , \qquad \kappa^{(\ell)}=-k^c \ell^a \nabla_a \ell_c \ .
\eea

\medskip

\noindent {\bf Definition 1.} {\em Let us introduce the following terminology:
\begin{itemize}
\item[i)] The IH $({\cal H}, [\ell^a])$ is {\em degenerate} iff $\kappa^{(\ell)}=0$.  
\item[ii)] The MOTS section ${\cal S}$ is outermost {\em stable} (in the $-k^a$ direction) 
if there exists a vector $X^a=\psi(-k^a)$, with $\psi$ a positive function, 
such that $\delta_X \theta^{(\ell)}\geq 0$. The section is {\em marginally stable}
if $\delta_X \theta^{(\ell)}= 0$ and {\em strictly stable} if
$\delta_X \theta^{(\ell)}\geq 0$ and $\delta_X \theta^{(\ell)}\neq 0$ somewhere.
\item[iii)] The  MOTS ${\cal S}$ is {\em future} iff $\theta^{(k)}\leq 0$, and 
{\em strictly future} if, in addition, $\theta^{(k)}\neq 0$ somewhere.  
\end{itemize}
}
For a discussion of the MOTS deformation operator $\delta_X$ along $X^a$, see \cite{AndMarSim,BooFai07}.

\medskip

\noindent{\bf Definition 2.} {\em Given the (spherical) surface ${\cal S}$, we will denote by
$\ell_o^a$ the rescaling of $\ell^a$ with divergence-free fundamental form $\Omega^{(\ell_o)}_a$
\bea
\label{e:Omega_div_free}
D^a \Omega^{(\ell_o)}_a = 0  \ .
\eea 
}
The existence of $\ell_o^a$ follows from the Hodge decomposition 
$\Omega^{(\ell)}_a = \epsilon_{ab} D^b \omega + D_a \lambda$ on a sphere. 
A rescaling $\psi>0$ on ${\cal S}$ (unique up to a multiplicative constant)
can be found
\bea
\label{e:rescaling}
\ell^a = {\psi} \cdot \ell_o^a \qquad , \qquad k^a = {\psi}^{-1} \cdot k_o^a \ .
\eea
Then, $\Omega^{(\ell_o)}_a \equiv  - k_o^c {q^d}_a \nabla_d (\ell_o)_c = \Omega^{(\ell)}_a - D_a\ln\psi$,
so that 
\bea
\label{e:omega_lambda}
\Omega^{(\ell_o)}_a= \epsilon_{ab} D^b \omega\qquad , \qquad \psi= \mathrm{const}\cdot e^\lambda \ .
\eea
MOTS stability characterizations are invariant under null normal rescalings
by a positive function $f$ \cite{Jaramillo:2011pg}:
$\ell^a \to f \cdot \ell^a$, $k^a \to f^{-1} \cdot k^a$, $\psi \to f \cdot \psi$.
Therefore in point {\em ii)} of Definition 1 we can substitute $\ell^a$ and $k^a$
by $\ell_o^a$ and $k_o^a$.

\medskip
Before stating the main result in Theorem 1, we revisit in Lemma 1 and 
Corollary 1 some known results in the literature. For the sake of a self-contained
presentation, we provide explicit proofs adapted to the horizon characterizations in Definition 1. 

\medskip

\noindent {\bf Lemma 1} \cite{Booth:2007wu,Mars:2012sb}. 
{\em
Given an axisymmetric IH and a section ${\cal S}$ adapted to axisymmetry, it holds 
\bea
\label{e:stability_kappa_theta}
\delta_{\psi(-k_o)} \theta^{(\ell_o)} =  -\kappa^{(\ell)} \theta^{(k_o)} \ .
\eea
}

\begin{proof}
First, on an IH  it holds \cite{Ashtekar:2001jb} 
\bea
\label{e:IH_conditions}
{\cal L}_\ell q_{ab} =0, \ \ 
{\cal L}_\ell \Omega^{(\ell)} = 0, \ \ 
\kappa^{(\ell)}=\mathrm{const}, \ \ [{\cal L}_\ell, D_a]=0, \ \ \delta_\ell \theta^{(k)} = 0\ .
\eea
Let us consider on ${\cal S}$ the null normals $\ell_o^a$ and $k_o^a$ in 
Eq.(\ref {e:rescaling}).
Then $\ell^a = \psi \ell_o^a $,  with $\psi > 0$ defined up to a factor not depending on ${\cal S}$.
From (\ref{e:IH_conditions}) we can choose $\psi$ with ${\cal L}_\ell \psi = 0$, so 
\bea
\label{e:l_properties}
\delta_\ell \theta^{(k_o)} = \delta_{\psi\ell_o} \theta^{(k_o)}=0 \ .
\eea
Second, let us denote the axial Killing on ${\cal S}$ as $\eta^a$, with 
${\cal L}_\eta q_{ab}={\cal L}_\eta \Omega^{(\ell)}_a=0$. Then, 
${\cal L}_\eta \psi = 0$ and ${\cal L}_\eta \omega = 0$ and
for any axisymmetric $A$
\bea
\label{e:deriv_Omega_dir}
\Omega^{(\ell_o)}_a  D^a A = \epsilon^{ab} D_b \omega D_a A = 0 \ .
\eea
Using this and $D^a \Omega^{(\ell_o)}_a = 0$, it follows \cite{BooFai07}
\bea
\label{e:MOTS_variation_expansion_axisym_lo}
\delta_{Ak_o} \theta^{(\ell_o)} &=&
 {}^2\!\Delta A  + A \left[ \Omega^{(\ell_o)}_a  {\Omega^{(\ell_o)}}^a 
 -\frac{1}{2}{}^2\!R + G_{ab}k_o^a\ell_o^b  \right] \nn \ ,  \\
\delta_{A \ell_o} \theta^{(k_o)} &=& -\kappa^{(A \ell_o)} \theta^{(k_o)} 
+ {}^2\!\Delta A + A \left[ \Omega^{(\ell_0)}_a  {\Omega^{(\ell_o)}}^a 
  -\frac{1}{2}{}^2\!R + G_{ab}k_o^a\ell_o^b  \right] \ ,
\eea
with  $\kappa^{(A \ell_o)} =-k_o^c (A \ell^a_o) \nabla_a (\ell_o)_c$ and $G_{ab}$ the Einstein tensor. 
Subtracting both equations
\bea
\label{e:relation_variations}
\delta_{A \ell_o} \theta^{(k_o)} =  -\kappa^{(A \ell_o)} \theta^{(k_o)} - \delta_{A(-k_o)} \theta^{(\ell_o)} 
\ .
\eea
Making $A=\psi$, using (\ref{e:l_properties}) and noting   that
$\kappa^{(\psi \ell_o)}=\kappa^{(\ell)}$ (since ${\cal L}_\ell\psi=0)$, we obtain (\ref{e:stability_kappa_theta}).
\end{proof}
We note that Lemma 1 follows from Corollary 2 in  \cite{Mars:2012sb} 
by making there $u_\ell=\psi^2$.

\medskip

\noindent {\bf Corollary 1.} (Booth \& Fairhurst, Mars) 
{\em Let us consider an IH containing a
strictly future axisymmetric section ${\cal S}$. Then
${\cal S}$ is marginally stable  iff $({\cal H}, [\ell^a])$ is degenerate.}
\begin{proof}
Marginal stability follows from degeneracy simply by  making
$\kappa^{(\ell)}= 0$ in (\ref{e:stability_kappa_theta}), 
without further assumptions. The reciprocal follows 
{\em ad absurdum} by assuming a non-vanishing (constant) $\kappa^{(\ell)}$,
using the  strictly future assumption and applying  (\ref{e:stability_kappa_theta}).

\end{proof}

Corollary 1  establishes \cite{Booth:2007wu,Mars:2012sb} the equivalence 
between marginal stability and degeneracy for strictly future MOTS. More generally,
in \cite{Booth:2007wu,Mars:2012sb} the stability/extremality of IHs containing a strictly 
future section ${\cal S}$ is classified by the sign of $\kappa^{(\ell)}$ so that, in particular, 
$\kappa^{(\ell)}\geq 0$ and MOTS stability are equivalent. Notably, proposition 3 in
\cite{Mars:2012sb} establishes such classification independently of the
topology of the horizon (with closed sections), for arbitrary dimension
and  without any axisymmetry assumption.

On the other hand, inequality $A\geq 8\pi |J|$ has been proved to hold
for stable axisymmetric MOTS \cite{Jaramillo:2011pg,Clement:2011tq}, where 
$A$ is the area of ${\cal S}$ and $J=1/(8\pi)\int_{\cal S} \Omega^{(\ell)}_a \eta^a dS$ is its (Komar) 
angular momentum. Furthermore, a rigidity result in terms of extreme Kerr sections
holds in the equality case \cite{Dain:2011pi,Jaramillo:2011pg,Mars:2012sb,GabJarRei12}.
The following theorem extends the equivalence in Corollary 1 to include the equality 
(rigidity) case $A = 8\pi |J|$ for IHs containing a strictly future section ${\cal S}$. 
Although the specifically new result in this theorem refers to such enlarged equivalence,
for the sake of a more clear and comprehensive presentation, we formulate it as a statement 
gathering known results with the new ones, in the spirit of providing a complementary counterpart
(valid for the degenerate case) of Corollary 5 in \cite{Mars:2012sb}
(that is focused on non-degenerate horizons).  

\medskip

\noindent {\bf Theorem 1.} {\em Let $({\cal H}, [\ell^a])$ be an axially symmetric IH
in a four-dimensional spacetime $(M, g_{ab})$  satisfying the 
dominant energy condition. Assume the non-negativity of $\kappa^{(\ell)}$, i.e. $\kappa^{(\ell)}\geq 0$, 
and that there exists a strictly future axisymmetric section ${\cal S}$. Then
\bea
\label{e:inequality}
A \geq 8\pi |J| \ ,
\eea
and equality occurs iff the following conditions hold:
\begin{itemize}

\item[(i)] The intrinsic geometry $q_{ab}$ is that of extreme Kerr.
\item[(ii)] The divergence-free part $\Omega_a^{(\ell_o)}$ 
of the 
normal fundamental form $\Omega_a^{(\ell)}$ is that of extreme Kerr. 
Moreover, $\psi$ in $\ell^a = \psi \ell_o$ is fixed
up to constant by the extreme Kerr geometry.
\item[(iii)] It holds $G_{ab} k^a \ell^b=0 $ on ${\cal H}$, with $k^a$ normal to  sections Lie-dragged
from ${\cal S}$ along $\ell^a$.
\item[(iv)] ${\cal S}$ is marginally stable or, equivalently, ${\cal H}$ is degenerate,
i.e. $\kappa^{(\ell)}=0$.
\end{itemize}
}

\begin{proof}
As commented above, under the hypothesis of a strictly future ${\cal S}$, 
$\kappa^{(\ell)}\geq 0$ is equivalent \cite{Booth:2007wu,Mars:2012sb}
to MOTS stability for ${\cal S}$. Explicitly, using 
$\kappa^{(\ell)} \theta^{(k_o)}\leq 0$ in (\ref{e:stability_kappa_theta}) we get
\bea
\label{e:S_stable}
\delta_{\psi(-k_o)} \theta^{(\ell_o)}\geq 0 \ .
\eea
From MOTS stability, inequality (\ref{e:inequality})
follows directly applying Lemma 1 in  \cite{Jaramillo:2011pg}. Our interest here 
is to improve the rigidity results in  \cite{Jaramillo:2011pg}.  
With this aim, we revisit the proof in \cite{Jaramillo:2011pg}, tracking specially the equality case. 
First, noting $D^a  \Omega^{(\ell_0)}_a=0$, we evaluate 
\bea
\label{e:step_1}
\!\!\!\!\!\!\!\!\!\!\!\!\!\!\! \!\!\!\!\!\!\!\!\!\!\!\!\!\!\!
\frac{1}{\psi}\delta_{\psi(-k_o)}\theta^{(\ell_o)}&=& 
- {}^2\!\Delta \mathrm{ln}\psi -   
|D\mathrm{ln}\psi|^2 + 2 \Omega^{(\ell_o)}_a  D^a\mathrm{ln}\psi
-\left[ |\Omega^{(\ell_o)}|^2 -\frac{1}{2}{}^2\!R + G_{ab}k_o^a\ell_o^b  \right] .
\eea
From  (\ref{e:deriv_Omega_dir}) and the axisymmetry of $\psi$, we have
$\Omega^{(\ell_o)}_a  D^a\mathrm{ln}\psi =0$. 
Introducing the projection of $\Omega^{(\ell)}_a$ along $\eta^a$, 
$\Omega^{(\eta)}_a \equiv \frac{1}{\eta}\eta^b  \Omega^{(\ell)}_b \eta_a$
with $\eta=\eta^a \eta_a$, from axisymmetry it follows $\Omega^{(\eta)}_a = \Omega^{(\ell_o)}_a$. 
Multiplying by an arbitrary $\alpha^2$,
using $k_o^a\ell_o^b=k^a\ell^b$
and integrating by parts 
 \bea
\label{e:step_2}
\!\!\!\!\!\!\!\!\!\!\!\!\!\!\! \!\!\!\!\!\!\!\!\!\!\!\!\!\!\!
\int_{\cal S} \frac{\alpha^2}{\psi}\delta_{\psi(-k_o)}\theta^{(\ell_o)}dS&=& 
\int_{\cal S}\alpha^2\left[ - |\Omega^{(\eta)}|^2 +\frac{1}{2}{}^2\!R - G_{ab}k^a\ell^b \right]dS \nn\\
&& + \int_{\cal S} \left[ 2D\alpha\cdot(\alpha D\mathrm{ln}\psi) - 
|\alpha D\mathrm{ln}\psi|^2\right]dS\nn \\
&\leq& \int_{\cal S}\alpha^2\left[  |\Omega^{(\eta)}|^2 +\frac{1}{2}{}^2\!R  \right]dS
-  \int_{\cal S} \alpha^2 G_{ab}k^a\ell^b dS +  \int_{\cal S} |D\alpha|^2dS \ ,
\eea
where we have used Young's inequality
\bea
|D\alpha|^2 \geq 2D\alpha\cdot(\alpha D\mathrm{ln}\psi) - 
|\alpha D\mathrm{ln}\psi|^2 \ ,
\eea
with equality iff $\alpha D\mathrm{ln}\psi = D\alpha$, that is iff $\psi = \mathrm{const} \cdot \alpha$.  
We can further write\footnote{Note that axisymmetry of $\alpha$ is not enforced. This recasts
Lemma 1 in \cite{Jaramillo:2011pg} and provides a closer link 
to the variational discussion of stable minimal surfaces \cite{Dain:2011pi}.}
\bea
\label{e:step_3}
\!\!\!\!\!\!\!\!\!\!\!\!\!\!\! \!\!\!\!\!\!\!\!\!\!\!\!\!\!\!
\int_{\cal S}\left[  |D\alpha|^2 +\frac{1}{2}\alpha^2 \;{}^2\!R \right]dS&\geq&
\int_{\cal S} \alpha^2|\Omega^{(\eta)}|^2 dS 
+ \int_{\cal S} \frac{\alpha^2}{\psi}\delta_{\psi(-k_o)}\theta^{(\ell)}dS
+  \int_{\cal S} \alpha^2 G_{ab}k^a\ell^b dS  \nn \\
&\geq& \int_{\cal S} \alpha^2|\Omega^{(\eta)}|^2 dS \ ,
\eea
where we have used the stability property (\ref{e:S_stable}) and the energy condition 
$G_{ab}k^a\ell^b \geq 0$. 
Equality happens iff:
\bea
\label{e:equality_stability}
\delta_{\psi(-k_o)}\theta^{(\ell_o)} = 0  \ \ \ , \ \ \ G_{ab}k^a\ell^b= 0 \ \ \ , \ \ \
\psi =  \mathrm{const} \cdot \alpha \ .
\eea
Inequality (\ref{e:step_3}) permits to match 
the reasoning in \cite{Jaramillo:2011pg}, leading to 
a  variational problem whose solution provides inequality (\ref{e:inequality}).
Equality occurs at the unique minimum of the
action functional and when conditions (\ref{e:equality_stability}) are fulfilled.
This happens iff:
\begin{enumerate}
\item[1.] The intrinsic geometry $q_{ab}$ is that of extreme Kerr
(this is proved in \cite{Dain:2011pi,Jaramillo:2011pg}; 
see also Corollary 5 in \cite{Mars:2012sb}).
Point {\em (i)} follows.

\item[2.] First, the divergence-free part of 
$\Omega^{(\ell)}_a$, i.e. $\Omega^{(\ell_o)}_a=\Omega^{(\eta)}_a$,
is fixed by the potential $\omega$  in (\ref{e:omega_lambda}) 
on an extreme Kerr section (this is proved in \cite{Dain:2011pi,Jaramillo:2011pg}).
Second, in the variational problem, the form of $\alpha$ is 
determined by $q_{ab}$ on ${\cal S}$ \cite{Dain:2011pi,Jaramillo:2011pg}. 
Therefore from {\em (i)}, at the unique minimum realizing 
equality in (\ref{e:inequality}), $\alpha$ is
determined by the intrinsic geometry of an  extreme Kerr section.
Using $\psi =  \mathrm{const} \cdot \alpha$ in  (\ref{e:equality_stability})
point {\em (ii)} is proved.

\item[3.] Point {\em (iii)} follows on ${\cal S}$ from  $G_{ab}k^a\ell^b= 0$ in  
(\ref{e:equality_stability}) (this is explicitly shown in \cite{Mars:2012sb}).
To extend it to the whole horizon, foliate ${\cal H}$ 
by Lie-dragging ${\cal S}$ along $\ell^a$. 
Each section of the foliation provides a uniquely defined $k^a$ and 
is a strictly future axisymmetric MOTS (due to the IH structure), so that the analysis on 
${\cal S}$ can be repeated on it.

\item[4.] The marginal stability of ${\cal S}$ follows from $\delta_{\psi(-k_o)}\theta^{(\ell_o)} = 0$
in  (\ref{e:equality_stability}). By Corollary 1 this is equivalent to
the degeneracy of ${\cal H}$, $\kappa^{(\ell)}=0$. This proves point  {\em (iv)}. 

\end{enumerate}

\end{proof}

\paragraph{Discussion.} Theorem 1 fills the following gaps in the literature:
\begin{itemize}
\item[a)] Refs.~\cite{Ansorg:2007fh,Hennig:2008zy} show 
that degeneracy in  (electro-)vacuum  axisymmetric Killing horizons 
implies  $A = 8\pi |J|$ (these results actually include charges~\footnote{See 
\cite{Lewandowski:2002ua,Kunduri:2011ii} for stronger rigidity results on the
geometry of degenerate local horizons.}). 
Theorem 1 recasts this result for axisymmetric IHs and, more interestingly, 
proves the reciprocal if the IH contains a strictly future section.
Following \cite{Clement:2011tq} this result extends straightforwardly
to the charged case, so that 
$(A/(4\pi))^2 \geq (2 J)^2 + (Q_{\mathrm{E}}^2 + Q_{\mathrm{M}}^2)^2$ (with $Q_{\mathrm{E}}$ and
$Q_{\mathrm{M}}$ the electric and magnetic charges, respectively) holds for IHs with
non-negative $\kappa^{(\ell)}$ containing a strictly future axisymmetric section.
Equality happens
iff the horizon is degenerate and satisfies points {\em (i)-(iii)} applied to Kerr-Newman.
This geometrises and proves the conjecture formulated in 
\cite{Ansorg:2007fh}~\footnote{Assuming horizon stability,
a similar result follows from the proof of Theorem 1 of \cite{Hennig:2008zy}. 
I thank C. Cederbaum for pointing this out~\cite{Ceder12}.}.

\item[b)] Inequality $A\geq 8\pi |J|$ is studied in \cite{Mars:2012sb} 
for non-degenerate stable IHs, showing that equality corresponds 
to marginal stability and the possibility of foliating  ${\cal H}$ by minimal surfaces.
Here we focus on the complementary degenerate case, so that Theorem 1 establishes the
conditions for the equivalence among $A = 8\pi |J|$, 
marginal stability and degeneracy for 
horizons containing a strictly future section. A weaker version of the minimal surface result 
in \cite{Mars:2012sb} follows from Lemma 1,
when dropping the future condition and imposing $\kappa^{(\ell)}\neq 0$ in Theorem 1.
The combined results in \cite{Mars:2012sb} and Theorem 1 here 
improve the rigidity analysis in \cite{Dain:2011pi,Jaramillo:2011pg}.

\item[c)] The result about $\psi$ in {\em (ii)} of Theorem 1 offers some insight on 
the function $\alpha$ in the variational problem, as a rescaling between null normals. 
It explains the following remark \cite{GabJarRei12}: on a section of extreme Kerr it holds 
$\ell^a_{\mathrm{K}} = \mathrm{const'}\cdot \alpha_{\mathrm{K}} \ell^a_o$, with
$\ell^a_{\mathrm{K}}$ the generator of ${\cal H}$ extending
to a Killing vector in extreme Kerr and $\alpha_{\mathrm{K}}$ 
the evaluation of $\alpha$ on extreme Kerr. It can
be interpreted  as stating that in the equality case of (\ref{e:inequality}), 
also the exact part of $\Omega^{(\ell)}_a$ is given by that in extreme Kerr.

\end{itemize}

\paragraph{Acknowledgments.} I am indebted to M.~E. Gabach-Cl\'ement and M. Mars for key insights.
I also thank M. Ansorg, C. Cederbaum, S. Dain, M. Reiris and W. Simon.


\section*{References}
\begin{thebibliography}{10}




\bibitem{Ashtekar:2001jb} 
 A.~Ashtekar, C.~Beetle and J.~Lewandowski,
  Class.\ Quant.\ Grav.\  {\bf 19}, 1195 (2002).


\bibitem{AndMarSim}
L.~Andersson, M.~Mars and W.~Simon,
\newblock Phys. Rev. Lett. {\bf 95}, 111102 (2005); Adv. Theor. Math. Phys. {\bf 12}, 853--888 (2008).


\bibitem{BooFai07}
I.~Booth and S.~Fairhurst,
\newblock Phys. Rev. {\bf D75}, 084019 (2007).

\bibitem{Jaramillo:2011pg}
J.~L. Jaramillo, M.~Reiris and S.~Dain,
\newblock Phys.Rev. {\bf D84}, 121503 (2011).

\bibitem{Booth:2007wu}
I.~Booth and S.~Fairhurst,
\newblock Phys. Rev. \textbf{D77}, 084,005 (2008).


\bibitem{Mars:2012sb}
 M.~Mars,
  Class.\ Quant.\ Grav.\  {\bf 29}  145019 (2012).


\bibitem{Dain:2011pi}
S. Dain and M. Reiris,
\newblock Phys.Rev.Lett. \textbf{107}, 051101 (2011).


\bibitem{GabJarRei12}
M.~E.~Gabach-Cl\'ement, J.L. Jaramillo \& M. Reiris, 2012; arXiv:1207.6761 [gr-qc]


\bibitem{Ansorg:2007fh}
M.~Ansorg and H.~Pfister,
\newblock Class.Quant.Grav. {\bf 25}, 035009 (2008).

\bibitem{Hennig:2008zy}
  J.~Hennig, C.~Cederbaum and M.~Ansorg,
  Commun.\ Math.\ Phys.\  {\bf 293}  449 (2010).


\bibitem{Lewandowski:2002ua}
  J.~Lewandowski and T.~Pawlowski,
  Class.\ Quant.\ Grav.\  {\bf 20}  587 (2003).

\bibitem{Kunduri:2011ii}
  H.~K.~Kunduri,
  Class.\ Quant.\ Grav.\  {\bf 28}  114010 (2011).

\bibitem{Clement:2011tq}
  M.~E.~Gabach-Cl\'ement and J.L.~Jaramillo, 2012 accepted in Phys. Rev. D;
  arXiv:1111.6248 [gr-qc].


\bibitem{Ceder12}
C. Cederbaum, {\em in preparation}.



\end{thebibliography}
\end{document}